\documentclass[fleqn,10pt]{article}

\usepackage{authblk}
\usepackage{graphicx}
\usepackage[a4paper, total={7in, 9in}]{geometry}
\usepackage{cite}
\usepackage[font={it}]{caption}
\title{Synaptic plasticity and neuronal refractory time cause scaling behaviour of neuronal avalanches}

\author[1,*]{L. Michiels van Kessenich}
\author[2,3]{ L. de Arcangelis}
\author[1]{ H. J. Herrmann}
\affil[1]{ETH Z\"urich,  Computational Physics for Engineering Materials, IfB}
\affil[2]{Dept. of Industrial and Information Engineering, Second University of Naples, Aversa (CE), Italy}
\affil[3]{INFN sez. Naples, Gr. Coll. Salerno (Italy)}
\affil[*]{laurensm@ifb.baug.ethz.ch}

\date{}

\begin{document}

\flushbottom
\maketitle

\begin{abstract}
Neuronal avalanches measured \textit{in vitro} and \textit{in vivo} in different cortical networks consistently exhibit power law behaviour for the size and duration distributions with exponents typical for a mean field self-organized branching process. These exponents are also recovered in neuronal network simulations implementing various neuronal dynamics on different network topologies. They can therefore be considered a very robust feature of spontaneous neuronal activity. Interestingly, this scaling behaviour is also observed on regular lattices in finite dimensions, which raises the question about the origin of the mean field behaviour observed experimentally. In this study we provide an answer to this open question by investigating the effect of activity dependent plasticity in combination with the neuronal refractory time in a neuronal network. Results show that the refractory time hinders backward avalanches forcing a directed propagation. Hebbian plastic adaptation plays the role of sculpting these directed avalanche patterns into the topology of the network slowly changing it into a branched structure where loops are marginal.
\end{abstract}

\thispagestyle{empty}

\section*{Introduction}

The scale-invariant behaviour of neuronal avalanches, reported by a large number of experiments \cite{plenz2014criticality,beggsPlenz2003,beggs2004neuronal,plenz2007organizing,shriki2013,pasquale2008self,mazzoni2007dynamics,petermann2009spontaneous,gireesh2008neuronal}, suggests that the brain is operating near a critical point. This conclusion is supported by the robustness of the scaling exponents measured experimentally {\it in vitro} and {\it in vivo} on different neuronal systems, from dissociated neurons to MEG on human patients. The exponent values are $1.5$ and $2.0$ for the avalanche size and duration distribution respectively, which characterize the mean field self-organized branching process \cite{zapperi1995}. Neuronal network models of integrate and fire neurons \cite{lucillaPRL,levina} inspired by self-organized criticality (SOC) \cite{BTW,pruessner} have been able to reproduce experimental observations, not only concerning the avalanche distributions \cite{lucillaBrainCrit}, but also the temporal organization of avalanches \cite{lombardiWaiting,Lombfrontiers}. Since in cortical systems it is difficult to measure directly the morphological synaptic connectivity, numerical studies have often implemented the structure of functional networks, as measured experimentally \cite{pajevic2009,massobrio2015,shefi2002morphological}. These networks have a complex structure \cite{makse} made of scale free modules \cite{cecchi}, connected by weaker links which make the entire network small world \cite{watts}. {\it In vitro} experiments on dissociated neurons \cite{pasquale2008self,mazzoni2007dynamics} indeed confirm that the synaptic network of connections has small world features, which could be the structural origin for the mean field behaviour of neuronal avalanches. However, studies on neuronal models have measured mean field exponents on a variety of networks, e.g. scale free, random, hierarchical and even regular networks in finite dimensions, like the square lattice \cite{lucillaActD}. This means that the observed exponents are surprisingly robust concerning changes in the underlying structure and different network topologies lead to the same universality class. This striking result seems counter-intuitive and raises the question of the origin of the mean field behaviour in neuronal avalanches. Previous studies \cite{moosavi2014,moosavi2015} have suggested that noise in the synaptic release can be a sufficient ingredient to obtain mean field exponents independent of the network. These results were obtained for the Zhang sandpile \cite{zhangModel}, where the noise was implemented as an additional random variable in the sand propagation for each toppling. However, this approach does not take into account neurobiological ingredients apart from integrate-and-fire elements. Furthermore, mean field exponents have also been found in neuronal models which do not feature noisy propagation \cite{lucillaActD,levina}, suggesting that the implemented neurobiological mechanisms originate the mean field behaviour.\\

In the present study we provide an answer to this open question by investigating the role of the structure of the supporting network and its relation with the neuronal activity evolving on it. The problem is addressed by considering the effect of some of the main characteristic ingredients of neuronal systems, in particular the neuronal refractory time and the plastic adaptation. We investigate how the interplay of these two phenomena, both of which act at the neuronal scale, lead to the global effect of changing the topology of the lattice into a branched structure where few loops are present.

\section*{Results}

Consider a neuronal network consisting of $N$ neurons connected by excitatory synapses, with weight $g_{ij}$. The initial network is a square lattice with periodic boundaries in the horizontal directions and open boundaries at top 
and bottom. The network is directed and therefore $g_{ij}$ is not necessarily equal to $g_{ji}$. The initial values for the synaptic weights can be either set all equal or randomly distributed and changing them does not influence the behaviour of our results. Each neuron is characterized by a membrane potential $v_i$ which can take values between zero and a threshold $v_c$. The initial value for the potentials is not important since the system will always evolve towards the critical state. Surpassing the threshold causes the generation of an action potential and the neuron fires. The action potential travels along the axon towards the synapses and then leads to a change in the potential of its connected neurons $j$ according to
\begin{equation}
v_{j}\rightarrow v_{j}+ v_{i}\frac{g_{ij}}{\sum_{k}g_{ik}}
\end{equation}
\begin{equation}
v_{i}\rightarrow 0 .
\end{equation}

Firing neurons can evoke further activity in connected neurons if their potential increases above the threshold $v_c$, which leads to the propagation of an avalanche. The firing rule is then applied until all neurons in the network are below the threshold and the configuration becomes stable. To trigger the next avalanche the system is driven by adding a small potential $\delta$ to random neurons. It is worth noticing that, if the firing rule is implemented on a network with all synaptic weights being equal, the model reduces to the Zhang sandpile \cite{zhangModel}, a well studied SOC model \cite{pruessner}. Two additional features characterizing real neurons are introduced in the model. Namely, the refractory time and activity dependent plasticity.\\

The firing rate of real neurons is limited by the refractory period, i.e., the brief period after the generation of an action potential during which a second action potential is impossible to elicit\cite{nicholls2001neuron}. In the model this is implemented by setting a neuron into a refractory state for $t_r$ time steps after it fired. During this period it does not receive stimulations from other neurons and will therefore not produce further action potentials. A firing neuron distributes its share of potential only to neighbours not in a refractory state. Therefore no dissipation is introduced in the potential distribution. The role of the refractory time on the avalanche propagation depends on the network considered. On a structure containing many back-and-forth connections, i.e. $i\rightarrow j$ and $j\rightarrow i$, a refractory time of one time-step has a very strong effect. If the topology contains very few of these connections then only a refractory time with $t_r>1$ will affect avalanche propagation significantly. The main consequence of the refractory time is that it hinders backwards avalanches, i.e. avalanche activities which activate regions already visited previously.
The propagation of each avalanche becomes directed away from the initial firing neuron which triggered the cascade, towards the boundary. This has some interesting consequences when combined with plastic adaptation.\\

Plastic adaptation governs how the structure of the network, namely the synaptic weights $g_{ij}$, evolve during the propagation of avalanches. It follows the principles of Hebbian plasticity, which strengthens the synapses between correlated neurons and weakens those which are not used \cite{hebb}. During the propagation of an avalanche connections between two successively active neurons are identified and strengthened proportionally to the potential variation
\begin{equation}
g_{ij}(t+1)=g_{ij}(t)+\alpha(v_{j}(t+1)-v_{j}(t))/v_c,
\end{equation}
where $\alpha$ is a parameter which determines the strength of plastic adaptation. It represents all possible biomolecular mechanisms which might influence the adaptation of synapses. After an avalanche comes to an end, all the weights in the network are weakened by the average strength increase:
\begin{equation}
\Delta g(t)=\frac{1}{N_b}\sum_{ij}\delta g_{ij}(t).
\end{equation}
If this weakening reduces the strength of a synapse below a minimum value, set here equal to $10^{-4}$, it is considered insignificant and the connection is removed from the network, a mechanism known as \textit{pruning}. The weakening occurring at all synapses depends on the total strength increase, leading to a constant sum of all synaptic weights. Furthermore, a maximum value $g_{max}$ for the weights is enforced which ensures that a steady state is reached, i.e. plastic adaptation saturates and the network becomes stable. The process of plastic adaptation in this model is governed by the two parameters, $\alpha$ and $g_{max}$. Raising $\alpha$ enhances the strength increase occurring at each firing event and therefore also accelerates the weakening and pruning of synapses. Thus, the parameter $\alpha$ is a measure for the speed of the plasticity process. At the end of the plasticity process most of the remaining synapses have a strength close to the maximal weight $g_{max}$. Increasing $g_{max}$ without changing the initial value for the weights implies that more bonds have to be weakened for another bond to reach the maximal value, since the sum $\sum g_{ij}$ stays constant. Therefore $g_{max}$ determines the percentage of pruned bonds once the plasticity process saturates.\\

\begin{figure*}[t]
    \centering
    \includegraphics[width=1\textwidth]{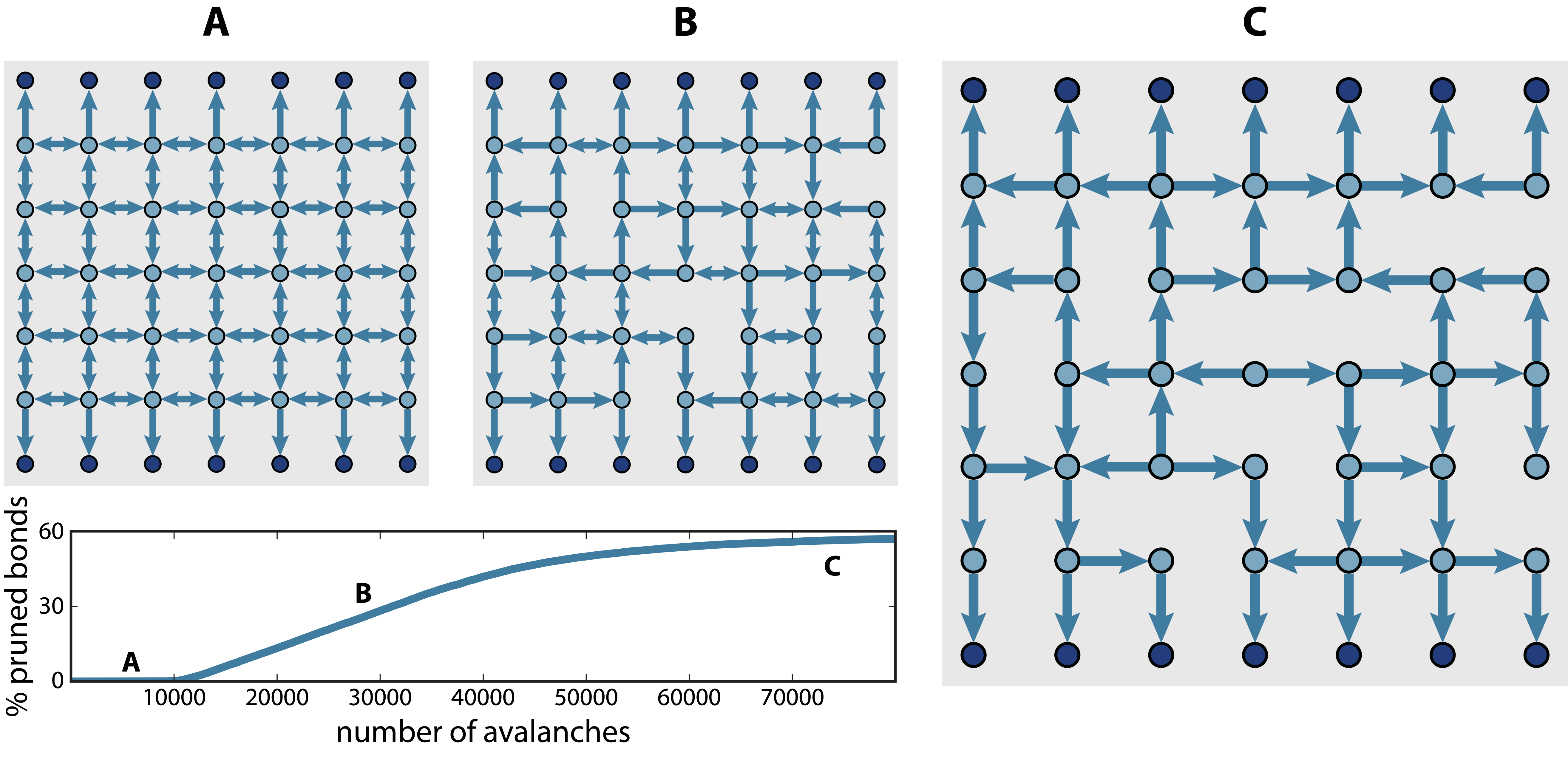}
	\caption{\textbf{Transition from the square lattice to a loop-less structure.} Three configurations at different stages during plastic adaptation and the corresponding pruning curve which shows the percentage of pruned bonds as a function of the number of avalanches. The initial structure is the square lattice (A). Each node represents a neuron and the arrows indicate whether a synapse is present between two neurons. The dark nodes constitute the open boundary at the top and bottom.  On the left and right are periodic boundary conditions. An intermediate state of the network is shown (B). The final configuration (C) is obtained when plasticity saturates. This final network has no loops, i.e. paths which start and end at the same node.}    
    \label{fig:exampleConfig}
\end{figure*}

Initially, the system is driven into a steady state by repeatedly triggering avalanches until the average potential of all neurons $\langle v\rangle$ becomes stationary. Then plastic adaptation is activated and the synaptic weights are modified according to the avalanche activity. Plastic adaptation imprints the shape of the avalanches onto the network. At the same time, the refractory time hinders backward avalanches and the propagation is directed away from the source neurons. As a consequence, synapses which would support backward avalanches get weakened and eventually pruned. Bonds which transport potential towards the boundary on the other hand are preferred and therefore strengthened. Figure \ref{fig:exampleConfig} illustrates the plastic adaptation process. Starting from a square lattice, plasticity sculpts the network according to the avalanche activity. The final configuration shows the network once the plasticity process saturates. Looking at the structure one notices that no loops are present in the network, i.e. paths that start and end at the same site. On the initial square lattice each pair of linked neurons has connections in both directions which facilitate backward avalanches. In order to understand why these back-and-forth connections get pruned, consider that neuron $i$ fires and triggers neuron $j$. The refractory time then forbids $j$ to fire back towards $i$. This leads to a strengthening of the connection $i\rightarrow j$ and a weakening of $j\rightarrow i$. During the following avalanches the probability for the propagation to flow in the same direction is higher as the weights have already been adjusted. Repeating this process during many avalanches will leave only one synapse surviving between $i$ and $j$. In general the refractory time hinders an avalanche to enter areas that were already active and therefore activity dependent plasticity weakens backward connections. The strength of this effect depends on the initial network and the length of the refractory period $t_r$. Therefore the combination of the two locally defined rules, plasticity and refractory time, have the global effect of removing loops from the network. The fact that the structures obtained have very few loops means that the avalanche propagation is transformed into a branching process. In the final configuration in figure \ref{fig:exampleConfig} one can observe that the branches might reunite and create structures reminiscent of anastomosis networks. These feature apparent loops. However, due to the directed nature of the connections, these do not allow for back-propagation. These structures are not forbidden by the refractory time and we do not consider them as loops as they do not create paths which start and end in the same site. The phenomenon of anastomosis is observed for example in leaf venation networks \cite{leafArchReview} or blood-vascular systems \cite{schaffer2006}. In these networks the direction of the propagation is not imposed by a refractory time but rather by a pressure gradient. It has been shown that anastomosis networks optimize resilience in the case of damage of connections\cite{katiforiDamage} and load fluctuations\cite{corson2010fluctuations}.\\

\begin{figure}[t]
    \centering
    \includegraphics[width=0.5\columnwidth]{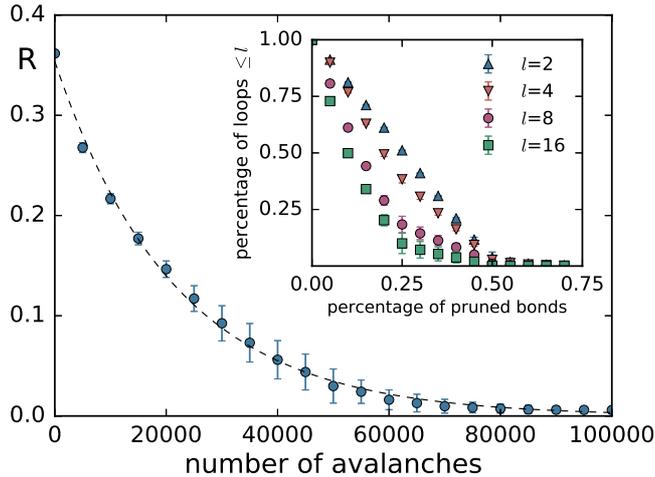}
        \caption{\textbf{The transition to a branching process can be illustrated by monitoring the flow of potential.} $R$ is the average fraction of potential variation at neurons which have previously been active during the propagation of one avalanche. In a branched network $R$ is zero. The plasticity process modifies the square lattice and the propagation of avalanches becomes a branching process. The dashed line shows an exponential fit. The inset shows the percentage of loops of length $\leq l$ remaining in the network at different stages of pruning. The number of loops goes to zero during plasticity and larger loops disappear more rapidly than smaller loops.}
    \label{fig:fractionOfFiringsFromYetSilentNeurons}
\end{figure}

To quantify the transition of the avalanche propagation towards a branching process we measure the sum of the potential variations induced in all neurons during one avalanche and calculate the fraction $R$ of these variations occurring in neurons which already fired in the same avalanche. Certainly, stimulating a neuron which has previously already been active requires a loop in the structure. In contrast, in a branching process the activity propagates away from the firing seed and cannot return to previously stimulated neurons. This implies that the fraction $R$ is zero for a branched network and greater than zero for networks with loops. Therefore we monitor $R$ as the original network is modified by plastic adaptation, to verify if the network evolves towards a branching network. Starting with the initial square lattice the structure is changed according to the plasticity rules, in the presence of the refractory time. After the network was modified during $N_{a}$ avalanches, plasticity is stopped and the avalanche distributions are measured. This measurement is done without the refractory time to ensure that the effect of all the loops in the network can be observed and the transition to a branched network becomes apparent. Figure \ref{fig:fractionOfFiringsFromYetSilentNeurons} shows how $R$ changes at different stages of plastic adaptation. On the initial square lattice, about 35\% of potential propagating through the system during one avalanche is directed towards neurons which were already active. Plastic adaptation clearly modifies the original network and the ratio $R$ decreases as the avalanche propagation turns into a branching process.
To understand how the underlying network is changing one can extract the number of loops remaining in the network using the elementary cycle algorithm by D. B. Johnson\cite{johnsonCycle}. To obtain the total number of loops of any length is computationally not feasible as the number of possible paths grows with the path length $l$ as $l!$. Therefore the search is limited to loops of length $l$, or shorter. The inset in figure \ref{fig:fractionOfFiringsFromYetSilentNeurons} shows the percentage of loops of length $\leq l$ remaining in the network at different stages of plasticity, i.e. different percentages of pruned bonds. The number of loops decreases during plasticity and goes to zero as pruning proceeds indicating that the network becomes branched. Furthermore one can see that larger loops disappear more rapidly than smaller loops.\\

\begin{figure}[t]
    \centering
    \includegraphics[width=0.5\columnwidth]{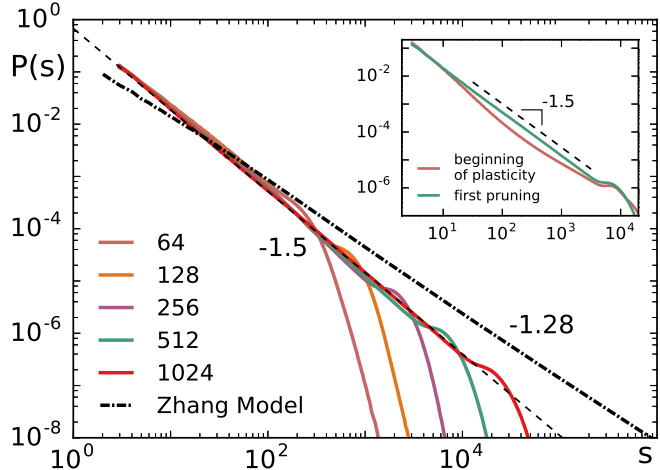}
	\caption{\textbf{Avalanche size distribution for various system sizes with refractory time and plastic adaptation.} The initial network is the square lattice which is modified by plasticity. The system exhibits critical scaling with an exponent of $1.5$. The parameters used are $t_r=1$, $\alpha=0.4$ and $g_{max}=2$. The dash-dotted line shows the size distribution for the Zhang model on a square lattice with a slope of $-1.28$. The dashed line shows the slope $-1.5$. The inset shows the distributions obtained with refractory time at the beginning of plastic adaptation and when the first bonds are pruned for $L=512$.}    
    \label{fig:avalancheAreaDist}
\end{figure}

After plasticity saturates we continue stimulating the network and measure the avalanche statistics. Since this neuronal model without refractory time and plastic adaptation is equivalent  to the Zhang sandpile, one might expect the same scaling exponent, i.e. $1.28$ on a two dimensional square lattice \cite{lubeck1997large}. Interestingly, the change in structure resulting from these two neurobiological mechanisms leads to a different exponent. Figure \ref{fig:avalancheAreaDist} shows the avalanche size distribution obtained when the plasticity process saturated. Networks with different system sizes exhibit a scaling behaviour with the exponent $1.53$, followed by an exponential cut-off controlled by the system size. The exponent was obtained with maximum likelihood estimation. A corresponding goodness of fit test was conducted using the Kolmogorov–Smirnov method\cite{clauset2009} resulting in a p-value of $0.1$. Moreover, the fact that the distribution scales with the system size indicates that the system is critical. This can be verified by measuring the branching ratio $\sigma$ which is defined as the average number of descendants per ancestor and only for $\sigma=1$ the system is in a critical state\cite{beggsPlenz2003,carvalhoSigma2000}. For the propagation of avalanches in this model we find $\sigma=1$ and the value of $\sigma$ does not change throughout plasticity. In Figure \ref{fig:avalancheAreaDist} we also show the size distribution obtained with the model in absence of plastic adaptation and refractory time on the square lattice, i.e. the Zhang model, which scales with the expected exponent $1.28$. We then investigate when the mean field scaling behaviour sets in during plastic adaptation. To achieve this avalanche statistics are measured at different stages of plastic adaptation. At the beginning of plastic adaptation the weights have only been slightly adjusted and many loops are still present. At this point the refractory time leads to a distribution with an initial slope steeper than $1.5$ (see inset in figure \ref{fig:avalancheAreaDist}).  Applying plastic adaptation until the very first pruning occurs, seems then sufficient to obtain scaling with an exponent 1.5. Although at this point loops remain in the network the weights of these loops are small and close to the pruning threshold and therefore do not contribute significantly to the avalanche propagation. Further plastic adaptation does not change the exponent.\\

\begin{figure*}[t]
    \centering
    \includegraphics[width=\textwidth]{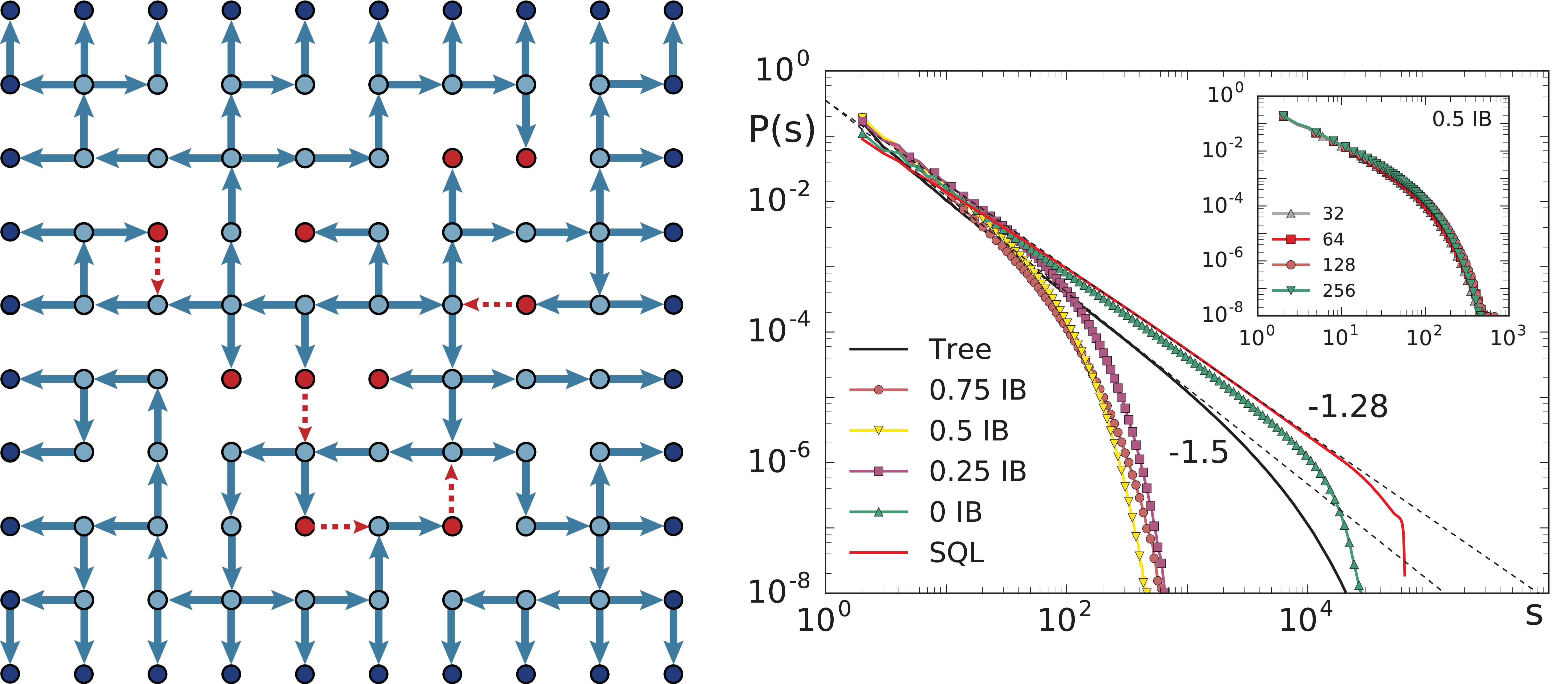}
	\caption{\textbf{Transition from a spanning tree to a square lattice by adding bonds.} On the left a spanning tree is shown (blue arrows). The dark blue nodes indicate the external boundary and the red nodes the internal boundary (IB). Initially bonds are added (red-dashed arrows) at the IB reducing the size of the IB and introducing loops into the network. In this configuration 50\% of the IB is removed by adding edges. On the right the avalanche size distributions obtained during this process are shown. On the spanning tree the distribution scales with an exponent of $1.5$. Completely removing the internal boundary is already sufficient to obtain an exponent close to $1.28$ and adding more bonds until the complete square lattice (SQL) is obtained does not change the exponent significantly. The inset shows that the distributions, obtained on a network where 50\% of the IB is removed, do not scale with the system size.}
    \label{fig:treeToSqlZhang}
\end{figure*}

To better understand the relevance of the topology of the network, we have analysed the problem also with an inverse approach. We start from a directed spanning tree on a square lattice and add bonds in order to monitor the transition towards the complete square lattice. More precisely, connections are added until all sites in the system are linked to their nearest neighbours and have an outgoing degree of four. The spanning tree is generated using the Wilson algorithm \cite{wilsonSpanningTree}, slightly modified to enforce that all sites on the outer edges of the square lattice have an outgoing degree $k_{out}=0$. Adding bonds to a spanning tree leads to the creation of loops. To study the effect of these loops, connections are added according to the following procedure: Each branch in a spanning tree ends with a site which has an outgoing degree of zero, $k_{out}=0$, and acts as the boundary. We distinguish two cases. The internal boundary (IB) are all the sites with $k_{out}=0$ inside of the square lattice. Whereas, the external boundary are all sites with $k_{out}=0$ on the edges of the square lattice. We start by adding bonds at the sites belonging to the IB. This is done to ensure that when approaching the square lattice it will only have a boundary on the outer edges and not within the structure. This process is illustrated in figure \ref{fig:treeToSqlZhang}.
In a spanning tree the average size of the IB is proportional to the total number of sites $N$, in contrast to the external boundary which grows proportionally to $\sqrt{N}$. Therefore, the IB does not vanish in the thermodynamic limit $N\rightarrow\infty$. Thus, not removing the IB when transitioning towards the square lattice leads to a break down of critical scaling at the intermediate stages. After dealing with the IB the procedure is continued by adding more bonds at random sites until the square lattice is completed. Since we start from a branching structure and add loops to the network this can be considered as an inverse approach compared to the removal of connections caused by plasticity and refractory time as described before. At different stages of this process we measure the avalanche size distribution and its exponent. To study the effect of loops, plastic adaptation and refractory time are not implemented as they would lead to a removal of the loops which were added. Interestingly, we find that just adding enough bonds to completely remove the internal boundary is sufficient to change the exponent from $1.5$ to $1.28$, as shown in Figure \ref{fig:treeToSqlZhang}. The structure is no longer a branching tree as it now contains loops but it is still distinct from the square lattice. Partially removing the internal boundary sites, e.g. removing only half of them, has an interesting effect on the distribution of avalanches. On a spanning tree it is always possible that during some avalanche the whole system becomes active. When adding a few bonds, and therefore loops, we observe that the avalanches do not reach the full system size anymore. This is illustrated in figure \ref{fig:treeToSqlZhang} where the distributions for intermediate stages fall off rapidly and avalanches are much smaller then the system size. Furthermore, the avalanches do not scale with the system size (see inset figure \ref{fig:treeToSqlZhang}). The reason why the largest avalanches become smaller when adding bonds to a spanning tree is discussed in the supplementary material. After the IB is completely removed, adding more bonds toward the square lattice does not change the exponent any further. The change in exponent from $1.5$ to $1.28$ is therefore rather abrupt and adding few bonds and therefore loops, whilst taking care of the IB adequately, is sufficient to obtain $1.28$. This is a further evidence for the relevance of loops for the avalanche exponents.

\section*{Discussion}

By means of numerical simulations of a neuronal network we show evidence that the origin of the mean field behaviour observed consistently for spontaneous avalanche activity is due to the structure of the supporting network where the presence of loops is marginal. Starting from a regular lattice, the loop-less structure is the outcome of two well established neurobiological mechanisms, namely neuron refractoriness and plastic adaptation. The combination of these two ingredients modifies the initial network, hampering back-propagation and selecting a structure without loops. This network transformation is at the origin of the change in scaling exponents from their value in finite dimensions to mean field values.  It is worth to note that the elimination of loops can also be obtained by spike-time-dependent plasticity \cite{kozloski} and would therefore lead to similar results when investigating the exponents of the avalanche distributions. An inverse approach, implemented by decorating a spanning tree to transform it into a regular network, confirms the crucial role of loops in the scaling behaviour of neuronal avalanche propagation. Please note that experimental results, unlike the presented model, show that spatially contiguous activation of nearest neighbours only accounts for about 40\% of avalanche activity \cite{beggsPlenz2003}. Due to the initial square lattice and the fact that connections are only pruned and not added, the present study allows exclusively nearest neighbour activations. This work aims at demonstrating why mean field exponents are obtained on different network topologies, even on regular lattices\cite{lucillaPRL}, and should not be taken as an accurate description of experiments. The same model has previously been implemented on other network topologies, including scale free and small world networks which represent a more accurate description of experimental results\cite{lucillaBrainCrit}.\\

This observation can be of more general interest for a variety of problems in biology where networks evolve dynamically and backward propagation is hindered. Any system which hinders backward flow in combination with an activity dependent plastic adaptation will remove loops. In our study the refractory time plays the role of forbidding backward propagation but in other systems the backward flow can be hindered by, for example, a pressure gradient. This is for instance the case in cardiovascular or leaf vasculature networks. These networks emerging due to the anastomosis mechanism contain loops but these loops have a direction and it is not possible to go back to a part of the network which has already been visited previously, much like the structures obtained in our study using the refractory time in combination with activity dependent plasticity.

\section*{Acknowledgements}

The Authors would like to thank D.R. Chialvo for his helpful contributions and a careful reading of our manuscript.\\
We acknowledge financial support from the European Research Council (ERC) Advanced Grant 319968-FlowCCS.

\section*{Author contributions statement}
L.M.v.K conducted the numerical simulations, L.d.A. and H.J.H conceived the research. All authors contributed to the writing of the manuscript.

\section*{Additional information}
\textbf{Competing financial interests}: The authors declare no competing financial interests.

\end{document}